\documentclass{preprint}

\usepackage{listings}
\usepackage{cite}
\usepackage{comment}
\usepackage{soul}
\usepackage{svg}
\usepackage{courier}
\usepackage{xurl}
\usepackage{float}
\usepackage{paralist}
\usepackage{tabularx}
\usepackage{booktabs}

\newfloat{lstfloat}{htbp}{lop}
\floatname{lstfloat}{Listing}

\newcommand{\qref}[1]{\href{#1}{Q\ref{#1}}}
\title{Extending and Applying Automated HERMES Software Publication Workflows} 
\author{Sophie Kernchen\autref{1}\autref{*}, Michael Meinel\autref{2}\autref{*}, Stephan Druskat\autref{3}\autref{*}, Michael Fritzsche\autref{4}, David Pape\autref{5}, Oliver Bertuch\autref{6}} 
\institute{
  \autlabel{1}\email{sophie.kernchen@dlr.de},
  \autlabel{2}\email{michael.meinel@dlr.de},
  \autlabel{3}\email{stephan.druskat@dlr.de}\\
  Institute of Software Technology\\
  German Aerospace Center, Berlin, Germany
  
  \autlabel{4}Andreas-Gymnasium, Berlin, Germany
  
  \autlabel{5}Helmholtz-Zentrum Dresden-Rossendorf, Dresden, Germany

  \autlabel{6}Forschungszentrum Jülich, Jülich, Germany
  
  \autlabel{*}\footnotesize{Authors contributed equally.}
}
\abstract{
    Research software is an important output of research and must be published
    according to the FAIR Principles for Research Software.
    This can be achieved by publishing software with metadata under a persistent
    identifier.
    HERMES is a tool that leverages continuous integration to automate
    the publication of software with rich metadata.
    In this work, we describe the HERMES workflow itself,
    and how to extend it to meet the needs of specific
    research software metadata or infrastructure.
    We introduce the HERMES plugin architecture 
    and provide the example of creating a new HERMES plugin 
    that harvests metadata
    from a metadata source in source code repositories.
    We show how to use HERMES as an end user,
    both via the command line interface,
    and as a step in a continuous integration pipeline.
    Finally, we report three informal case studies whose results provide a preliminary evaluation 
    of the feasibility and applicability of HERMES workflows, and the extensibility of the \texttt{hermes} software package.
} 

\begin{document}
\maketitle
    
\section{Introduction}
An important goal that every Research Software Engineer (RSE) should follow, is to make software compliant with the FAIR Principles for Research Software~\cite{ChueHongEtAl2022}. 
In order to make software FAIR, it should be published with rich metadata
in a way that supports its identification through persistent identifiers
such as DOIs.
The publication of software is also appropriate due to its
status as an important product of research~\cite{JayEtAl2021}. 
Software published in the way described above
furthermore allows for formal software citation~\cite{smithSoftwareCitationPrinciples2016},
which in turn supports the reproducibility of research results
that have been gained using the software.

Software publication, i.e. the deposition of software artifacts to open access
repositories that provide persistent identifiers (PIDs), 
such as \href{https://inveniosoftware.org/products/rdm/}{InvenioRDM}\footnote{\url{https://inveniosoftware.org/products/rdm/}}, \href{https://dataverse.org/}{Dataverse}\footnote{\url{https://dataverse.org/}}, or similar, 
can be automated to a large degree.
Currently, however, the addition, maintenance, curation and publication 
of software metadata specifically is often a manual process\cite{concept-paper}.

The project HERMES\cite{meinel_hermes} (see~\autoref{sec:hermes}) automates the process of publishing software with rich metadata\cite{concept-paper}.
This paper presents HERMES software publication workflows and their extension mechanism via plugins for the \texttt{hermes} software package.
As a preliminary evaluation of the applicability and extensibility of HERMES workflows, 
we report results and experiences from a live coding workshop conducted by the authors,
as well as informal case studies for applying HERMES in source code repositories,
and extending HERMES with a new plugin. In these preliminary evaluations, we focus on the following questions 
that inform the future development of HERMES:

\begin{enumerate}[Q1.]
    \item How effectively does the HERMES concept described in~\cite{concept-paper} enable users 
    to publish software with rich metadata, and what are implementation challenges? \label{rq-concept-implementation}
    \item How effectively do the \texttt{hermes} package and its documentation enable
    software developers without prior knowledge about HERMES publication workflows to extend the \texttt{hermes} package with a new plugin? \label{rq-extensibility}
    \item How can the \texttt{hermes} package and its documentation be improved to support different stakeholder needs? \label{rq-improvement}
\end{enumerate}

\section{HERMES}\label{sec:hermes}
HERMES (\textit{HElmholtz Rich MEtadata Software Publication}) is an open source project initially funded by the \href{https://helmholtz-metadaten.de/de}{Helmholtz Metadata Collaboration}\footnote{\url{https://helmholtz-metadaten.de/de}} under the grant ZT-I-PF-3-006.
The HERMES tools help users automate the publication of rich metadata together with their software projects and each of its versions.
They can automatically harvest and process software metadata, and submit them to tool-based curation, approval and reporting processes.
Software versions can be deposited on publication repositories that provide PIDs (e.g. DOIs).

\subsection{HERMES project}\label{subsec:project}
The central outputs of the HERMES project are a concept for automating workflows for software publication with rich metadata\cite{concept-paper}, a Python software package called \texttt{hermes}\cite{meinel_hermes},
and templates that enable \texttt{hermes}' use in different continuous integration systems.
\texttt{hermes} implements the single phases of the workflow while the HERMES workflow itself is configured in continuous integration instructions.

The adoption of the continuous integration configuration templates 
provided by the project allows
RSEs to focus more on their actual work.
HERMES provides \href{https://docs.software-metadata.pub/en/latest/}{documentation}\footnote{\url{https://docs.software-metadata.pub/en/latest/}} for \texttt{hermes}' \href{https://docs.software-metadata.pub/en/latest/api/index.html}{Python API}\footnote{\url{https://docs.software-metadata.pub/en/latest/api/index.html}}, 
and \href{https://docs.software-metadata.pub/en/latest/tutorials/automated-publication-with-ci.html}{tutorials to get started}\footnote{\url{https://docs.software-metadata.pub/en/latest/tutorials/automated-publication-with-ci.html}}.
The Python API has been designed to enable interoperability across
many different metadata types and infrastructure ecosystems.
In configuring a HERMES publication workflow,
RSEs decide which environments (e.g., source code repository platforms, metadata sources,
publication repositories) to use.
Additionally, the data model allows RSEs to use HERMES as a framework 
and build additional automated processes.

HERMES aims to build an open community,
where everyone is welcome to contribute.
The extensibility of the tools developed within the HERMES community
makes them adoptable for the needs of their users.

\subsection{HERMES workflow}\label{subsec:tool}
The HERMES workflow is designed to be run within a continuous integration (CI) pipeline independent of the CI infrastructure in use.
Supported platforms include, but are not limited to, \href{https://docs.github.com/en/actions}{GitHub Actions}\footnote{\url{https://docs.github.com/en/actions}} and \href{https://docs.gitlab.com/ee/ci/}{GitLab CI}\footnote{\url{https://docs.gitlab.com/ee/ci/}}.
The pipeline-based approach enables a push-based model, compared to other pull-based workflows (e.g. the \href{https://developers.zenodo.org/#update-schedule}{Zenodo-GitHub integration}\footnote{\url{https://developers.zenodo.org/\#update-schedule}})~\cite{concept-paper}.
Running \texttt{hermes} on in-house resources, e.g., your own source code repository platform instance and continuous integration system, also reduces the dependency on third-party services.

\begin{figure}[h!]
		\centering
		\includegraphics[width=15cm, keepaspectratio]{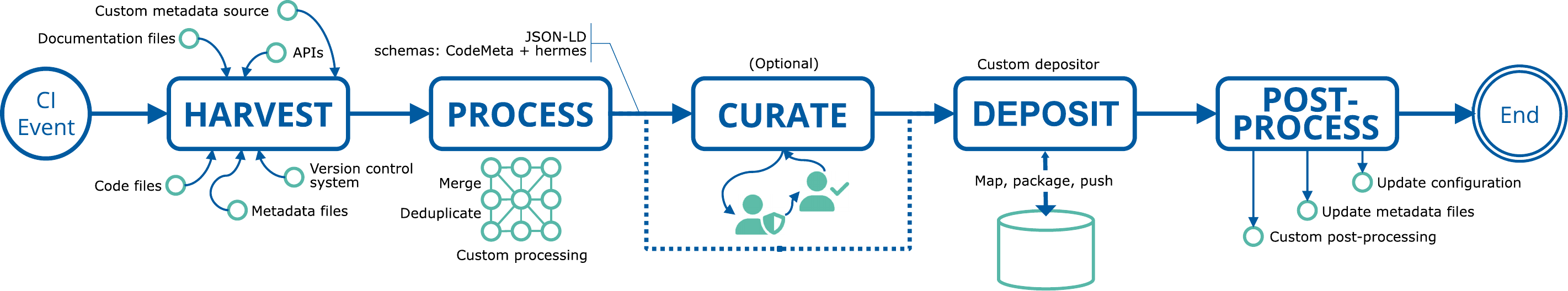}
        \caption{The five phases that form the HERMES workflow: \textit{harvest}, \textit{process}, \textit{curate}, \textit{deposit}, and \textit{postprocess}}
		\label{fig:work}
\end{figure}

The overall HERMES workflow consists of five phases that form a pipeline as shown in \autoref{fig:work}.
In the pipeline, each phase needs to run successfully before the subsequent phase can be started.
Each phase of the workflow has a clear objective:


\begin{description}\label{desc:steps}
    \item [harvest]
    During this phase, metadata is collected into a common, well-defined data model that is derived from CodeMeta~\cite{JonesEtAl2023}.
    As rich descriptive metadata is the key element to useful software publication, \texttt{hermes} strives to collect all metadata from source code repositories and connected platforms.
    It supports the collection from structured metadata sources like Citation File Format~\cite{druskat2021CFF} files, CodeMeta files, or Git history.
    However, it can also be extended to retrieve unstructured metadata, e.g., from a README file or a CONTRIBUTORS list.
    
    \item [process]
     After the \textit{harvest} phase collects all retrievable metadata and stores them in separate files for each metadata source,
     these collected artifacts are then collated during the \textit{process} phase to produce a single set of metadata.
     The result is a consistent set of metadata in an extended CodeMeta format, i.e., a well-defined JSON-LD format.
     Additional metadata about the processed metadata tracks metadata provenance, i.e., the source of each metadata point.
     Where appropriate, we collaborate with other consortia to identify metadata standards that can be applied to metadata that cannot currently be represented as pure CodeMeta.
     
    \item [curate]
    The objective of this optional phase is to retain, manually check and validate the metadata that is to be published.
    The default implementation of this phase is a non-operation, i.e., the collated metadata is copied to the output without changes.
    During a given curation process implementation, the collated metadata may be improved according to users' requirements.
    Both the default non-operation, and any custom curation implementations, signal \texttt{hermes} that the data has been curated
    by writing output.
    An example of how the non-operation can be used for merge-based curation on GitHub or GitLab is implemented in the respective templates as described in \autoref{subsect:ci_github}.
    Further possible review methods are currently being investigated in the project \href{https://helmholtz-metadaten.de/de/inf-projects/softwarecard}{Software Curation and Reporting Dashboard (Software~CaRD)}\footnote{\url{https://helmholtz-metadaten.de/de/inf-projects/softwarecard}}.

    \item [deposit]
    After \textit{processing} and optional \textit{curation}, the resulting metadata files are \textit{deposited} in a publication repository.
    This deposition target can be a long-term storage that provides a persistent identifier (PID) and access to the metadata, a software registry or directory solution that makes the software better findable, a local file system, or anything in between.
    A deposition to different targets is also possible by running this phase repeatedly with different configurations within the same workflow run.
    
    \item [post-process]
    The last phase provides a means to clean up and feed back information from the workflow to the original (source code) repository, or targets that are not publication platforms.
    This can include, but is not limited to, backtracking changes from the publication repository (e.g., by storing a newly minted PID in the structured metadata in the source code repository).
\end{description}

All phases can be configured individually and can also be extended with new functionality using a plugin interface.
Each phase is run independently\footnote{Of course this is only true as long as the process does not need output data from a previous phase.} as a separate process.
It is also possible to invoke a certain phase several times.
This flexibility allows HERMES to be adapted to multiple use cases, for example in mandatory organizational processes.
A collection of pre-defined workflows for common processes is available to help users get started.

The next section explains the extension mechanism in general and takes a closer look at how to provide a new \texttt{hermes} harvesting plugin.

\section{Extending HERMES}
The \texttt{hermes} Python package can be used and adapted in different ways.
As \texttt{hermes} is published open source under an Apache 2.0 license, users can fork it and adapt the software to their needs.
\texttt{hermes} also defines a public API, so that the package can be integrated into other tools as a library.
This approach is taken, for example, in the Software~CaRD project.
However, the primary intended use case is to use the command line interface and configuration file to integrate \texttt{hermes} into CI templates, standalone software, or build scripts.

Early in the HERMES research project's (see~\autoref{subsec:project}) funding phase, we ran a stakeholder workshop\footnote{\url{https://events.hifis.net/event/205/}} to gather and discuss requirements for an automated workflow for software publication with rich metadata.
As a result of this workshop and further ad-hoc community consultations, we were able to define requirements, the two most important in the context of this paper being:

\begin{enumerate}
    \item the workflow uses formats and schemas that follow \textbf{community standards};
    \item any software implementing the workflow is \textbf{extensible} to enable its application for combinations of different metadata sources, different infrastructure components, different disciplinary requirements, and different organizational requirements.
\end{enumerate}

We implement this in \texttt{hermes} by using the community standard schema for descriptive software metadata CodeMeta~\cite{JonesEtAl2023}
as universal exchange format between workflow phases, and as output format.
CodeMeta's serialization as JSON-LD furthermore allows us to extend it to represent additional harvested metadata.
We also harvest CodeMeta files, as well as the community standard format for software citation metadata, the Citation File Format~\cite{druskat2021CFF}.

Extensibility is implemented in a plugin architecture that uses the standard Python entry point definition~\cite{PythonEntryPoints}.
The \texttt{hermes} API also exposes a plugin interface that can be used to extend \texttt{hermes} without modification of the core source code.
The following section explains the plugin structure and provides the necessary know-how to build a new plugin.

\subsection{Plugin architecture}\label{subsec:struct}

\begin{figure}[h!]
		\centering
		\includegraphics[width=.8\textwidth, keepaspectratio]{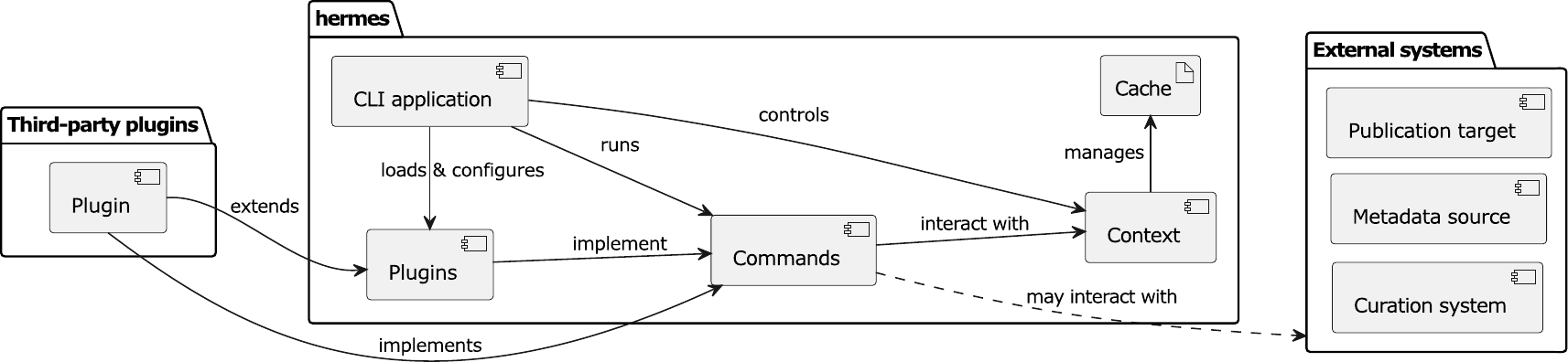}
		\caption{\texttt{hermes} components and interactions with external components.}
		\label{fig:comps}
\end{figure}

\texttt{hermes} provides a plugin architecture for third parties to contribute additional features.
The basic structure is the same for every plugin.
There is a generic base class \textit{HermesPlugin} that provides basic functionality:

\begin{itemize}
    \item The execution context that was used to invoke the plugin.
    This functionality is encapsulated in the \textit{HermesCommand} class and requires adapting the specific behaviour according to the way \texttt{hermes} was executed.
    This interface is also used to access the \texttt{hermes} configuration.

    \item Methods to load and store data from the \texttt{hermes} caches.
    This allows access to the results of previous phases as well as provision of the current plugin's results to other plugins for further processing.

    \item A possibility to define additional, plugin specific settings.
    Configuration options are implemented using \texttt{pydantic-settings}~\cite{ColvinEtAl2023}.
    This allows to declaratively define new settings options and validate the settings upon loading.
    It also supports different sources for settings out of the box (e.g., reading from a configuration file, loading from environment variables).
\end{itemize}

\begin{figure}[h!]
		\centering
		\includegraphics[width=.7\textwidth, keepaspectratio]{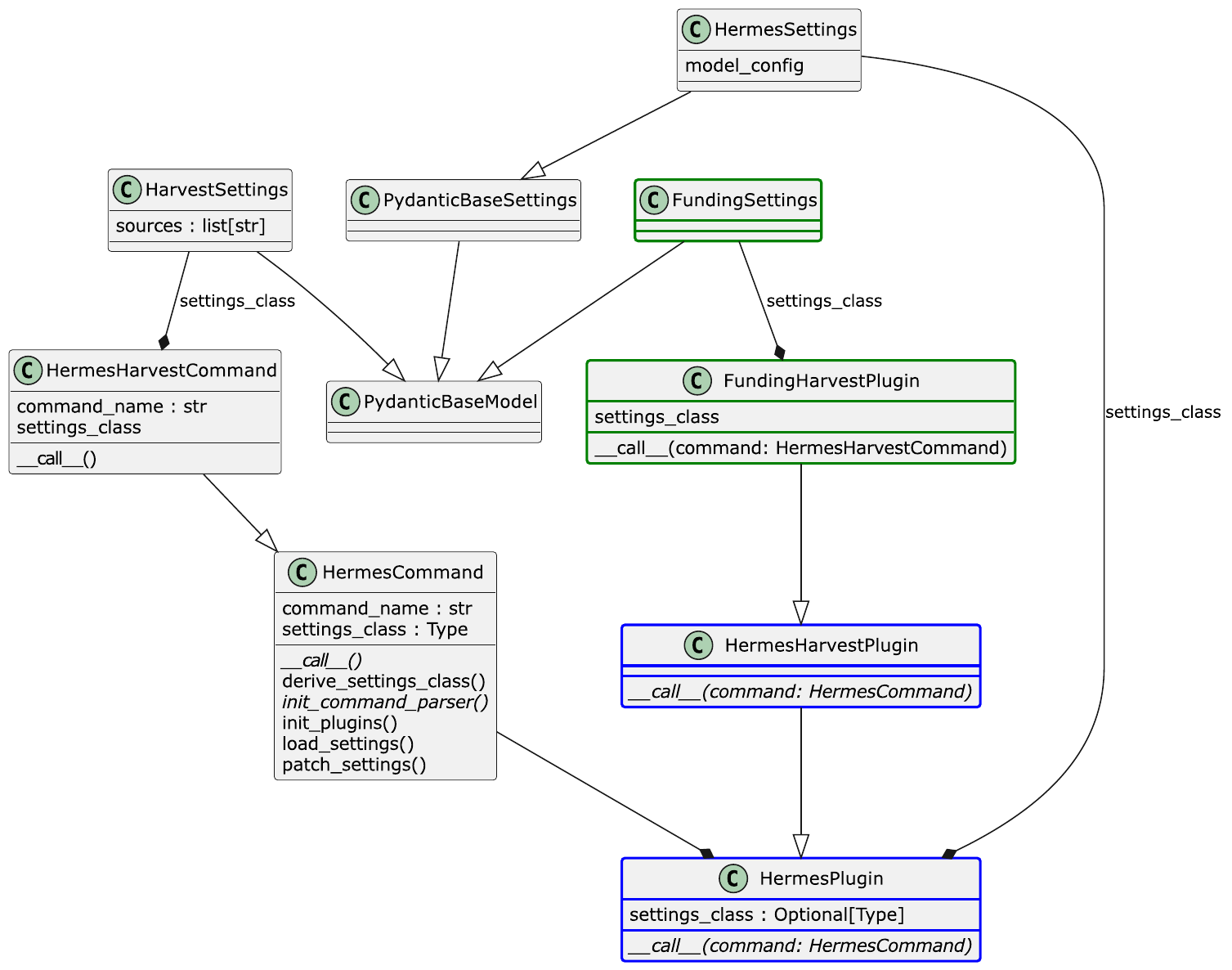}
		\caption{Partial class structure for a new harvest plugin \texttt{FundingHarvestPlugin} showing the classes that implement the new plugin (green border) and parent classes (blue border).}
		\label{fig:plug_struc}
\end{figure}

For each phase, there is a derived base class available that adds the basic processing logic that is required.
In this \textit{HermesPlugin} class there is one abstract method \textit{\_\_call\_\_ } which is overwritten by the derived classes to reflect the basic flow.
\autoref{fig:plug_struc} shows a reduced UML class diagram including an exemplary implementation of a new plugin that inherits from a base class.
\texttt{hermes} provides some built-in plugins for typical basic HERMES workflows:

\begin{description}
    \item [git] A \textit{harvest} plugin that collects all contributors and committers from the Git history.
    \item [cff] A \textit{harvest} plugin that collects citation metadata found in a \texttt{CITATION.cff}~\cite{druskat2021CFF} file.
    \item [codemeta] A \textit{harvest} plugin that collects metadata from an existing CodeMeta~\cite{JonesEtAl2023} file.
    \item [invenio deposit, invenio\_rdm deposit] Two \textit{deposit} plugins that publish finalized metadata in Invenio~\cite{RaoEtAl2021} or InvenioRDM~\cite{TheInvenioRDMteam} instances respectively.
    \item [file] A \textit{deposit} plugin that writes an extended CodeMeta file with all collected metadata.
    \item [invenio postprocess, invenio\_rdm postprocess] Two \textit{postprocess} plugins that provide functionality to: 
    \begin{inparaenum}[(1)]
        \item store the new record ID that was created during the deposition to Invenio(RDM) in the \texttt{hermes.toml} configuration file (see \autoref{sec:usage});
        \item record a new DOI in the \texttt{CITATION.cff} citation file.
    \end{inparaenum}
\end{description}

\subsection{How to write a plugin}\label{subsect:plugin_howto}

\texttt{hermes} exposes a number of Python extension points\cite{PythonEntryPoints} that are used to hook in new functionality.
There is one extension point for each phase of the workflow.
The extension point groups are structured as follows:

\begin{description}
    \item [hermes.harvest]
    Extend the \textit{harvest} phase with a new metadata source.
    The entry point needs to be a subclass of \textit{HermesHarvestPlugin} and override the \textit{\_\_call\_\_} method accordingly.
    
    \item [hermes.process]
    Extend the \textit{process} phase.
    This entry point is not well defined or finalized yet.
    However, the base class to derive from is \textit{HermesProcessPlugin}.
    
    \item [hermes.curate]
    Add new possible review methods to the \textit{curate} phase.
    A successful curation means that the output of this phase does not differ from the input.
    The base class that needs to be extended for this entry point is \textit{HermesCuratePlugin}.
    
    \item [hermes.deposit]
    Add new deposition targets to the \textit{deposit} phase.
    Currently, the \textit{HermesDepositPlugin} provides a base workflow that relies on the internal state of the deposit plugin.
    Internally, it follows a sub-workflow that takes care of

    \begin{itemize}
        \item projecting the metadata onto a supported schema,
        \item creating a container / entry to upload metadata (and optionally software artifacts) to, or selecting an existing one to be extended with a new version,
        \item uploading the metadata and software artifacts, and
        \item publishing the new deposition.
    \end{itemize}
    
    \item [hermes.postprocess]
    Extend the \textit{post-process} phase to feed back output from the HERMES workflow, e.g., into the repository.
    The base class for this entry point is \textit{HermesPostprocessPlugin}.
    
\end{description}

To write a new plugin, it is important to follow the given plugin structure (see \autoref{subsec:struct}).
\autoref{lst:plug} shows the structure of an example plugin that provides a new source for the harvest phase, in this example implementing harvesting of funding metadata as set in \texttt{hermes.toml} directly.
\autoref{fig:plug_struc} shows that \textit{FundingHarvestPlugin} inherits from the \textit{HermesHarvestPlugin} and uses a settings class \textit{FundingSettings} which is derived from Pydantic's \textit{BaseModel}.

The plugin's functionality is implemented in the \textit{\_\_call\_\_} method that is overridden by the plugin.
The method is meant to return a tuple of two dictionaries.
The first dictionary should contain the collected metadata as a CodeMeta dataset.
The second dictionary may contain additional meta-metadata that is not restricted to some schema (even though this is encouraged).
The additional meta-metadata can later be used, e.g., to keep a record of the metadata provenance or to help decide automatically how to collate different metadata sets.

Our example adds a new configuration key \textit{harvest.funding.grant\_id} to the workflow configuration file \texttt{hermes.toml} (see \autoref{sec:usage}).
If set to be non-empty, this value will be added as \textit{funding} metadata value.
The extension of the metadata is reflected by setting the meta-metadata value for \textit{added} accordingly.

\begin{lstfloat}[h]
\begin{lstlisting}[caption=Example for a plugin named "funding" that implements a new source for the harvest phase, label=lst:plug, language=Python]
from pydantic import BaseModel

from hermes.commands.harvest.base import HermesHarvestPlugin


class FundingSettings(BaseModel):
    grant_id: str = ""


class FundingHarvestPlugin(HermesHarvestPlugin):
    settings_class = FundingSettings

    def __call__(self, command):
        codemeta_data = {}
        plugin_metadata = {}

        funding_info = command.settings.funding
        if funding_info.grant_id:
            codemeta_data["funding"] = funding_info.grant_id
            plugin_metadata["added"] = ["funding"]

        return codemeta_data, plugin_metadata
\end{lstlisting}
\end{lstfloat}

\section{Using HERMES}
\label{sec:usage}
There are various ways to use HERMES. 
This section describes the usage of HERMES from the command line and within a continuous integration workflow, using the example of GitHub Actions. 
Note that this has been tested with Python 3.10 and \texttt{hermes} 0.8.0~\cite{meinel_hermes}, whose API must still be considered unstable.
We provide further information and up-to-date tutorials in the \href{https://docs.software-metadata.pub/en/latest/index.html}{HERMES Documentation}\footnote{\url{https://docs.software-metadata.pub/en/latest/index.html}}.

As a primary source of configuration information, \texttt{hermes} uses the file \textit{hermes.toml} in TOML format\footnote{\url{https://toml.io/}}.
This file contains information for all phases, each in a separate section named after the phase.

Most phases provide generic configuration options.
For example, the user can select which harvest plugins to activate and which deposition target(s) to use.
An overview of those generic options is listed in \autoref{tab:herm_toml}.

Plugins are referenced by their name, must be part of the respective entry point group, and implement the matching base class as described in \autoref{subsect:plugin_howto}.
The special settings that each plugin might introduce are collected in a subsection below the respective phase section.
I.e., the settings of the Citation File Format~\cite{druskat2021CFF} harvest plugin (named \textit{cff}) are collected in the section \textit{harvest.cff}.

\begin{table}[h!]
    \centering
    \caption{HERMES TOML configurations.}
    \label{tab:herm_toml}
    \resizebox{\textwidth}{!}{%
    \begin{tabular}{llll}
        \toprule
        Plugin & Setting & Type & Options\\
        \midrule
        \textit{harvest} & sources & List & Plugin name (e.g.,  "cff", "codemeta", "git") \\
        \textit{harvest.cff} & enable\_validation & Bool & Whether to validate \texttt{CITATION.cff} \\
        \textit{harvest.git} & branch & String & Name of branch that will be harvested \\
        \textit{deposit} & target & String & Plugin name (e.g., "invenio", "invenio\_rdm", "file") \\
        \textit{deposit.file} & filename & String & Path of file to deposit to \\
        \textit{deposit.invenio(\_rdm)} &  site\_url & String & URL of publication repository \\
        & communities & List & Communities to publish software into \\
        & access\_right & String & Access rights ("open", "embargoed", "restricted", "closed") \\
        & embargo\_date & String & Embargo date for software \\
        & access\_conditions & String & Conditions for software access \\
        & api\_paths &  Dict & API paths for keywords \\
        & auth\_token &  String & Token for target platform \\
        & files &  List[pathlib.Path] & Files/paths to deposit \\
        & record\_id & Int & Identifier of publication to append new version to \\
        & doi  &  String & DOI of software \\
        \textit{postprocess} & execute &  List & Tasks to execute \\
        \bottomrule
    \end{tabular}
    }
\end{table}

As \texttt{hermes} uses \texttt{pydantic-settings}, it is possible to set configuration options using environment variables.
The implementation does also allow to override configuration options using command line parameters.

\subsection{Using HERMES via command line}\label{subsec:cl}
\texttt{hermes} is published on the \href{https://pypi.org}{Python Package Index}\footnote{\url{https://pypi.org}}.
The command \texttt{pip install hermes} installs the latest version of \texttt{hermes}.
After installation, users can use \texttt{hermes} from the command line.
\texttt{hermes} comes with a single top-level command that provides sub-commands for each phase and some additional utility functions.
\autoref{tab:herm_com} shows all sub-commands and their functionality.

\begin{table}[!h]
    \centering
    \caption{An overview of the different sub-commands that the \texttt{hermes} command line interface provides~\cite{meinel_hermes}.}
    \label{tab:herm_com}
    \begin{tabular}{ll}
        \toprule
        Command & Description\\
        \midrule
            \texttt{help} &  Show help page and exit.\\
            \texttt{clean} & Clean up caches from previous \texttt{hermes} runs. \\
            \texttt{harvest} &  Harvest metadata from configured sources.\\
            \texttt{process} &  Process the collected metadata into a common dataset.\\
            \texttt{curate} &  Curate the unified metadata before deposition.\\
            \texttt{deposit} &   Deposit the curated metadata to repositories.\\
            \texttt{postprocess} &  Post-process the published metadata after deposition.\\
        \bottomrule
    \end{tabular}
\end{table}

For \texttt{hermes} to work, it is important to run the phases in the order as described in \autoref{subsec:tool}.
To enforce this, \texttt{hermes} keeps track of the phases that have already been run in an internal cache.
Each phase can be run repeatedly as long as the caches from the previous phases are valid.

\subsection{Using HERMES via continuous integration in GitHub}\label{subsect:ci_github}
HERMES was developed with application in continuous integration environments in mind.
As part of the project, we provide different templates in a separate \href{https://github.com/softwarepub/ci-templates/}{project on GitHub}\footnote{\url{https://github.com/softwarepub/ci-templates/}}.
Continuous integration in \href{https://github.com/}{GitHub}\footnote{\url{https://github.com/}} is configured using GitHub Actions.
Each action is described in a single YAML file located in the \texttt{.github/workflows/} directory of the respective Git repository.
Users can copy the matching HERMES template into the Git repository and adapt it accordingly.
To help with the adaptation, the templates have areas marked with \texttt{\#ADAPT} to highlight the contents that need to be changed.

The default GitHub template provides a set of jobs that use \texttt{hermes} for a pull-request-based curated deposition to \href{https://zenodo.org/}{Zenodo}\footnote{\url{https://zenodo.org/}} (or any other platform that runs InvenioRDM).
In addition to the basic HERMES workflow and the correct invocation of the different phases, it also offers an example of how the curation phase could be implemented.

For the \textit{deposition} phase, the user needs to provide a valid authentication token for the target platform, stored as a ``GitHub Secret''.
During the continuous integration run, the \textit{curation} phase opens a pull request to let users review the collected metadata.
To enable this, the user has to allow GitHub Actions to create pull requests.
Only when the curation pull request is successfully merged, the continuous integration workflow will continue to carry out the actual deposition in the target repository.
Note that all branches that are created for the curation are only temporary and will be deleted once the deposition was successful.
The final \textit{post-process} phase will again open a pull request contributing metadata that can be added back into the repository (e.g., a newly created PID for the latest deposition).
If the branch is not changed in the template, the HERMES workflow will run with every push to the \textit{main} branch.

\section{Preliminary evaluation}

After describing the concept for automated software publication workflows~\cite{concept-paper},
gathering requirements,
and implementing a proof-of-concept version of \texttt{hermes},
we took some first steps to evaluate these early outcomes.
Specifically, we wanted to test 
the feasibility of the HERMES workflow concept for developers of research software as its users (\qref{rq-concept-implementation}) and
the practical extensibility of the \texttt{hermes} package (\qref{rq-extensibility}), and
identify key areas for improvement of the \texttt{hermes} package and its documentation for both, users and plugin developers (\qref{rq-improvement}).

To test this and achieve a preliminary evaluation,
we ran a live coding workshop and conducted two informal case studies. 
We report on the preliminary results in the following sections.

\subsection{Case study 1: live coding workshop at deRSE24}
\label{subsec:workshop}

We ran a 90-minute hands-on live coding workshop at \textit{deRSE24 - Conference for Research Software Engineering in Germany} (\textit{Automating your FAIR software publications with HERMES – a hands-on workshop}) to evaluate the general
feasibility of the HERMES workflow approach detailed in~\cite{concept-paper}.

During the workshop, participants learned to use the \texttt{hermes} package~\cite{hermes_0_8_0} on the command line, as well as within a continuous integration workflow using Github Actions.
The audience consisted of 18 people.
Three participants had some prior knowledge about HERMES and its usage, and joined a breakout group where they were given a demonstration about the plugin mechanism of \texttt{hermes}.




The remaining 15 participants did not have any knowledge of HERMES, but at least basic knowledge of Python.
They were introduced to HERMES with a presentation.
After that a live coding session demonstrated HERMES, which allowed participants to follow along in real time.
The live coding included setting up a mock project, and installing and using \texttt{hermes} from the command line.
After successful set up for local use, we did an interactive walk through of the \href{https://docs.software-metadata.pub/en/latest/tutorials/automated-publication-with-ci.html}{HERMES tutorial ``Set up automatic software publishing''}\footnote{\url{https://docs.software-metadata.pub/en/latest/tutorials/automated-publication-with-ci.html}}.
After live coding, 8 participants had succeeded in setting up and using HERMES to publish their own software on Zenodo Sandbox.

For those participants that did not successfully complete the publication process with HERMES, 
we interactively elicited and recorded the reasons for failure during the workshop:

\begin{enumerate}
    \item Inability to install the \texttt{hermes} package due to either the lack of a Python installation on participants' machines, or failure to install the required Python version ($\geq$~3.10).
    \item Non-existent accounts required to follow the live coding using the respective platforms (GitHub, Zenodo Sandbox), and therefore inability to follow the live coding due to delays while creating accounts.
\end{enumerate}

Both reasons show failure of the workshop organizers to provide clear technical requirements
to participants and assistance in their fulfillment ahead of the workshop,
and failure to check their fulfillment during
the workshop before live coding.
These failures are of methodological nature and do not
as such attest a lack of feasibility of applying HERMES workflows for automated software publication with rich metadata,
and thus cannot contribute to answering \qref{rq-concept-implementation} or \qref{rq-improvement}.

Beyond these issues, the workshop showed the general usability of HERMES,
and that it is possible for a principal stakeholder group -- RSEs and developers of research software -- to apply HERMES.
To answer our question \qref{rq-concept-implementation} above, we therefore partially conclude that HERMES' approach~\cite{concept-paper} is generally viable,
and that the current development status of \texttt{hermes} and its documentation
makes HERMES workflows fundamentally usable.
Beyond that, we were able to elicit first-hand user feedback
towards answering our question \qref{rq-improvement}:

\begin{enumerate}[\itshape{Issue} 1]
    \item \label{issue1} Using the default configuration, \texttt{hermes}~v0.8.0~\cite{hermes_0_8_0} provides very limited feedback to users.
    Logging output is only available in a hidden file,
    or by adapting the logging configuration.
    Errors coming from plugins are not reported clearly enough by \texttt{hermes}.
    This makes it hard to debug HERMES workflows.
    \item The \texttt{hermes} user interface should be improved to support inexperienced users in applying the
    automated workflow, i.e., adjust templates, supply credentials for infrastructure components (source code repository platform, publication repository platform).
\end{enumerate}

We addressed \textit{Issue 1} in \texttt{hermes} 0.8.1~\cite{hermes_0_8_1} by adapting the default logging configuration.
The log file is now output to the working directory and therefore more visible.
Also, more logging output is now directed towards the standard output.
This is especially helpful when HERMES is run in a CI context, and files produced during the process are not easily retrievable.
We also made sure that errors originating from plugins are output and better visible.

We furthermore address \textit{Issue 2} by introducing a new \texttt{hermes init} command that provides a dialogue-based guided setup of the HERMES workflow from the command line. It has been merged into the codebase\footnote{See \url{https://archive.softwareheritage.org/swh:1:rev:dc0f62585d7696ed1ec8380887811b3c661ec6c3}.} and will be released with version \texttt{0.9.0}.

We argue that having a locally installed Python version $\geq$~3.10 -- as proved to be an issue for some workshop participants -- is not necessary for the main use case
of HERMES workflows, where \texttt{hermes} is run remotely in a continuous integration environment and technical requirements are solved by defining a CI workflow
as per the HERMES documentation.
Overall, we found that a hands-on workshop as described above was useful for 
eliciting user feedback, as well as advocacy for automated software publication with HERMES.


\subsection{Case study 2: publish \texttt{hermes} on Zenodo using HERMES}
\label{subsec:dogfooding}

To further explore potential answers to our questions \qref{rq-concept-implementation} and \qref{rq-improvement},
we dogfooded \texttt{hermes} in a HERMES workflow
to publish the package on Zenodo with metadata harvested from different sources.

To achieve this, we followed our own tutorial to publish \texttt{hermes} version 0.8.1b1 on Zenodo\footnote{\href{https://zenodo.org}{https://zenodo.org}},
an instance of the InvenioRDM repository software~\cite{TheInvenioRDMteam} that is supported by HERMES as a publication target.
As collaborative development of \texttt{hermes} is using GitHub, 
we implemented the HERMES workflow using GitHub Actions.

The respective HERMES workflow made use of the optional curation phase (see~\autoref{subsec:tool}),
in which the processed metadata was made available before publication in a pull request\footnote{\href{https://github.com/softwarepub/hermes/pull/267}{https://github.com/softwarepub/hermes/pull/267}} against the
source code repository from which the software was being published.
Once the pull request was accepted, the publication process continued as expected,
successfully publishing \texttt{hermes} via HERMES~\cite{hermes_0_8_1b1}.
This successful dogfooding experiment provides positive preliminary evidence towards answering \qref{rq-concept-implementation},
while in the process also providing preliminary evidence towards \qref{rq-improvement} as described in the following.

The applied GitHub Action template provided in our documentation automates publication using a prescribed process.
If you follow the template closely, the publication process works out-of-the-box.
However, any specific requirements on the publication workflow require that the template is adapted,
which in turn requires an understanding of how the workflow configuration works.

An example of this is a change of publication trigger:
In our experiment, the publication of a new version should be triggered by creating a new Git tag for the version to be published,
and pushing that tag to the GitHub repository.
However, the template at experiment time included the example of a trigger event upon push to a specific branch.

We solved this by adjusting the workflow configuration to use a different Git command to create the Git branch which presents the processed software metadata for curation\footnote{\href{https://github.com/softwarepub/hermes/commit/d42aa8f41e507718e9b23d11825911aa4096c9aa}{https://github.com/softwarepub/hermes/commit/d42aa8f41e507718e9b23d11825911aa4096c9aa}}.
While this resulted in a successful publication on Zenodo,
the post-processing phase was not completed successfully.
The cause for this problem lies in GitHub internals:
When being triggered by a Git tag, it is not easily possible do determine the base branch from within a GitHub Action.
There are solutions available but they can not be easily integrated into the template in its current form.
Therefore, in answer to \qref{rq-improvement}, a more complex templating system may be needed to better support curation-based HERMES workflows.

Another pointer to the need for improved error logging (see also \autoref{subsec:workshop})
is insufficient feedback when supplying a wrong authentication token for the target publication platform.
During the experiment, we failed to change the token in the workflow configuration when we switched from
Zenodo Sandbox to Zenodo proper as the target publication platform.

We partially conclude that while the results of this dogfooding experiment
yield positive answers for \qref{rq-concept-implementation},
its more complex application case contributes to an answer for \qref{rq-improvement}.
Specifically, it provides further evidence for the need for better error logging and feedback (see~\href{issue1}{\textit{Issue~\ref{issue1}}} above),
and points to issues when configuring HERMES workflows that include the optional curation phase.
While this may not be the default use case, the existing issues need to be solved to provide full support
in \texttt{hermes}.
An alternative process and tooling for software metadata curation is currently investigated in the project Software~CaRD,
which is conducted with participation from the HERMES team.

\subsection{Case study 3: developing a harvester for \texttt{pyproject.toml}}
\label{subsec:plugin}
To explore potential answers to \qref{rq-extensibility}, we conducted another informal case study
involving the development of a new HERMES plugin by a less experienced Python developer.

In July 2024, a student intern (author Fritzsche) joined the
Institute of Software Technology of the German Aerospace Center for three weeks.
With his consent, he was tasked to develop a \texttt{hermes} plugin
that harvested metadata from Python \texttt{pyproject.toml} manifest files\footnote{See PEP 621 (\href{https://peps.python.org/pep-0621/}{https://peps.python.org/pep-0621/}).}.

The intern was 16 years old and attended 11th grade in a German secondary school when he started his internship.
According to himself, he had basic experience in programming with Python, but ``felt more comfortable'' programming in C\#.
He also could not prove any detailed knowledge of relevant Python specific standards, such as entry points or \texttt{pyproject.toml} files,
and had no prior knowledge about the HERMES project, software metadata, the FAIR principles, or the CodeMeta standard.
Our aim in conducting the informal case study was to gather evidence towards answering \qref{rq-extensibility}, 
and evaluating the level of proficiency required to develop a harvesting plugin for \texttt{hermes}.
We assume that proficiency in Python, as well as detailed knowledge about software metadata and software publication processes,
cannot be expected across the complete RSE population, 
which made this study especially interesting to evaluate the usability of the \texttt{hermes} plugin mechanism.
The intern is not an RSE and has never published software or written a \texttt{hermes} plugin before.
He was given a general introduction to HERMES during onboarding to this task -- comparable to the introductory talk during the deRSE24 workshop (see~\autoref{subsec:workshop}) -- and supervised by a \texttt{hermes} core developer 
This developer did not actively intervene with the intern's implementation work,
but was available to answer any questions by the intern.

To facilitate development, a fork of the \href{https://github.com/softwarepub/hermes-plugin-git}{\texttt{hermes} Git plugin}\footnote{\url{https://github.com/softwarepub/hermes-plugin-git}} 
was created and used as baseline, along with the \href{https://docs.software-metadata.pub/en/latest/tutorials/writing-a-plugin-for-hermes.html}{tutorial that describes the \texttt{hermes} Git plugin}\footnote{\url{https://docs.software-metadata.pub/en/latest/tutorials/writing-a-plugin-for-hermes.html}}.
The intern researched any missing information himself on the internet.
In the process, the intern rarely needed to query an expert (i.e., his supervisor).
Within two weeks of development time, the intern finished a working plugin that could be used with the \href{https://pypi.org/project/hermes/0.8.0/}{hermes 0.8.0}\footnote{\url{https://pypi.org/project/hermes/0.8.0/}} release~\cite{hermes_0_8_0}.
This development time included a longer period for setting up and configuring build automation for publishing the new plugin to PyPI.
To demonstrate that the metadata extracted from the \texttt{pyproject.toml} was complete,
we adapted a HERMES workflow to publish the new plugin to Zenodo~\cite{fritzsche_hermes_plugin}.

After the intern handed in the initial version of the plugin,
we solved minor technical issues, 
such as backporting the plugin to run on Python 3.10, 
and harvesting missing license metadata.
Such changes would usually be requested by maintainers of a software project during
a request to merge new code into an existing codebase.
In this case, we refrained from implementing a more complex request-based merge workflow due the the limited availability of the intern.
We take the small number of necessary changes to the intern's initial implementation to suggest an overall good maintainability of the plugin.
Overall, we take the intern's success to produce the plugin within a two week timeframe
as a pointer towards a positive potential answer to \qref{rq-extensibility}.

In terms of preliminary answers to our question \qref{rq-improvement}, we identified some issues during the experiment.
First of all, the existing Git plugin is not a good starting point for implementing new harvesters.
This is due to the fact that it includes internal processing of harvested data, such as matching different contributors, and differentiation between authors and contributors.
This is unnecessary code that artificially bloats a plugin meant to be used as a template for newly developed plugins.

Additionally, we found opportunities for improvement of the \texttt{hermes} data model,
which is both complex by nature and unsophisticated by design.
The expected data for internal processing for \texttt{hermes} is a mapping with CodeMeta-compliant linked datasets.
While this is already explicitly used in some of \texttt{hermes}' API, 
there is no library yet that allows the performance of common tasks on the dataset, 
such as matching two typed dictionaries, or merging two typed datasets.
This makes new plugin implementations more cumbersome and verbose than necessary.

\section{Conclusion and outlook}
We presented HERMES, an implementation of the
HERMES concept for automated software publication with rich metadata~\cite{concept-paper},
based on the \texttt{hermes} Python package~\cite{meinel_hermes}.
We also presented an extension mechanism to customize HERMES workflows for particular and highly customized use cases,
based on \texttt{hermes} plugins.
Such use cases include the harvesting of additional metadata sources,
or publishing to different target publication repositories.

We furthermore reported three informal case studies that we conducted to provide preliminary answers
to our questions \qref{rq-concept-implementation}, \qref{rq-extensibility} and \qref{rq-improvement}.
While being informal, the presented studies are valuable
in that they provide initial evidence from examining typical uses of HERMES workflows and the \texttt{hermes} package
by relevant stakeholders: RSEs and developers of research software as \textit{users} of HERMES workflows,
and as \textit{developers} of \texttt{hermes} plugins.

Two of the case studies (study 1 in \autoref{subsec:workshop} and study 2 in \autoref{subsec:dogfooding}) demonstrate that
the HERMES concept for automating software publication is feasible and fundamentally applicable 
with the current development status of the \texttt{hermes} package (\qref{rq-concept-implementation}).
Study 3 presented in \autoref{subsec:plugin} demonstrates that it is fundamentally possible
to extend \texttt{hermes} with new plugins.
In this case, the developer that we studied is not in fact part of either of the user target groups of 
research software engineers and researchers who code.
Nevertheless, we argue that the prior experience by the developer likely matches that of
researchers with some, if little, experience in writing code.
The study is therefore relevant, in that
it provides initial evidence that
extending \texttt{hermes} with plugins that provide new functionality is generally possible with adequate and realistic resources (\qref{rq-extensibility}).
All studies furthermore helped us identify future work to improve HERMES' applicability and extensibility
by pointing to some underlying problems that need to be solved in future work,
e.g., identifying usage patterns for workflows and deducting sensible defaults,
and determining configuration requirements for workflows (\qref{rq-improvement}).

Our preliminary studies exhibit clear limitations.
As case studies, they suffer from inherent shortcomings with regard to their sample size, reproducibility, generalization, internal validity, etc.
Additionally, the reported cases were not selected in a fully controlled fashion.
While we selected participants for study 1 from the population of participants in the workshop,
excluding those with prior (practical) knowledge of HERMES,
we did not actively control for this or other variables.
Case study 2 is also limited with regard to potential researcher bias,
as it was conducted with participants -- the HERMES developers and project team -- 
who have extensive and detailed insights into the project, including the concept and implementation of HERMES workflows.
Case study 3 is potentially limited by the role of the studied developer (secondary school student),
which does not reflect the primary target group for HERMES.
As argued above, we believe that regardless of this fact,
this limitation is less severe as to our knowledge,
there are no factors that qualify high school students more than
academic researchers with regard to research software development.

Future work should build on the preliminary evidence presented here
to apply more formal research methods to evaluate the HERMES workflow concept
and the relevant technical artifacts.
More precisely, we believe that a design science methodology for software engineering,
e.g., as put forward by Wieringa~\cite{wieringa_2014_design_science_methodology},
should be applied. In this context, the HERMES workflows would represent conceptual and methodological, and the \texttt{hermes}
package and its documentation technical, artifacts, whose development could be treated as an \textit{improvement problem}.

As a first step in this direction, HERMES' context should be methodically re-evaluated
in terms of the stakeholders to consider.
While our informal user studies cover important conceptual and technical questions,
they focus on only two of at least three stakeholder groups: RSEs and developers of research software,
and developers of \texttt{hermes} plugins.
A third stakeholder group should be taken into account going forward: research software infrastructure providers.
Given the heterogeneous landscape of infrastructure components in use at research organizations -- for example,
different source code repository platforms, different publication repository platforms,
different organizational prerequisites, and different methods to create and maintain knowledge of software publication --
future research projects on software publication should work with infrastructure providers such as 
libraries and computing centers;
More precisely, the feasibility, adoptability and usability of HERMES workflows should be investigated
in the context of different infrastructure ecosystems and across different research disciplines.

HERMES is still under active development by the original developers.
Future technical work within the HERMES project includes the development of additional plugins
to support harvesting metadata sources, deposition and post-processing targets,
as well as API design and documentation work to make the project more accessible
to the wider community.

\begin{acknowledge}
The HERMES project (ZT-I-PF-3-006) was funded by the Initiative and Networking Fund of the \href{https://www.helmholtz.de/en/about-us/structure-and-governance/initiating-and-networking}{Helmholtz Association}\footnote{\url{https://www.helmholtz.de/en/about-us/structure-and-governance/initiating-and-networking}} in the framework of the \href{https://helmholtz-metadaten.de/de}{Helmholtz Metadata Collaboration’s}\footnote{\url{https://helmholtz-metadaten.de/de}} \href{https://helmholtz-metadaten.de/en/inf-projects}{2020 project call}\footnote{\url{https://helmholtz-metadaten.de/en/inf-projects}}.
The authors thank their collaborators at the Helmholtz-Zentrum Dresden-Rossendorf (Dr. Oliver Knodel, Guido Juckeland, Tobias Huste) and Forschungszentrum J\"ulich (Nitai Heeb).
The authors also thank three anonymous reviewers for the time and effort they afforded.
Their constructive comments contributed greatly to improving this paper.
\end{acknowledge}



\bibliography{article}

\newcommand{\etalchar}[1]{$^{#1}$}
\begin{thebibliography}{MDK{\etalchar{+}}24d}

\bibitem[CJR23]{ColvinEtAl2023}
S.~Colvin, E.~Jolibois, H.~Ramezani.
\texttt{pydantic-settings} ({{Version}} 2.1.0).
\emph{PyPI}, Nov. 2023.
\\\url{https://pypi.org/project/pydantic-settings/2.1.0/}

\bibitem[CKB{\etalchar{+}}22]{ChueHongEtAl2022}
N.~P. Chue~Hong, D.~S. Katz, M.~Barker, A.-L. Lamprecht, C.~Martinez, F.~E. Psomopoulos, J.~Harrow, L.~J. Castro, M.~Gruenpeter, P.~A. Martinez, T.~Honeyman, A.~Struck, A.~Lee, A.~Loewe, B.~{van Werkhoven}, C.~Jones, D.~Garijo, E.~Plomp, F.~Genova, H.~Shanahan, J.~Leng, M.~Hellstr{\"o}m, M.~Sandstr{\"o}m, M.~Sinha, M.~Kuzak, P.~Herterich, Q.~Zhang, S.~Islam, S.-A. Sansone, T.~Pollard, U.~D. Atmojo, A.~Williams, A.~Czerniak, A.~Niehues, A.~C. Fouilloux, B.~Desinghu, C.~Goble, C.~Richard, C.~Gray, C.~Erdmann, D.~N{\"u}st, D.~Tartarini, E.~Ranguelova, H.~Anzt, I.~Todorov, J.~McNally, J.~Moldon, J.~Burnett, J.~{Garrido-S{\'a}nchez}, K.~Belhajjame, L.~Sesink, L.~Hwang, M.~R. {Tovani-Palone}, M.~D. Wilkinson, M.~Servillat, M.~Liffers, M.~Fox, N.~Miljkovi{\'c}, N.~Lynch, P.~Martinez~Lavanchy, S.~Gesing, S.~Stevens, S.~Martinez~Cuesta, S.~Peroni, S.~{Soiland-Reyes}, T.~Bakker, T.~Rabemanantsoa, V.~Sochat, Y.~Yehudi, R.~F. WG.
{{FAIR Principles}} for {{Research Software}} ({{FAIR4RS Principles}}) (1.0).
\emph{Research Data Alliance}, May 2022.
\\\doi{10.15497/RDA00068}

\bibitem[DBJ{\etalchar{+}}22]{concept-paper}
S.~Druskat, O.~Bertuch, G.~Juckeland, O.~Knodel, T.~Schlauch.
Software publications with rich metadata: state of the art, automated workflows and {HERMES} concept.
\emph{ArXiv} abs/2201.09015, 2022.
\\\doi{10.48550/2201.09015}

\bibitem[DSC{\etalchar{+}}21]{druskat2021CFF}
S.~Druskat, J.~H. Spaaks, N.~Chue~Hong, R.~Haines, J.~Baker, S.~Bliven, E.~Willighagen, {P{\'e}rez-Su{\'a}rez, David}, O.~Konovalov.
Citation {{File Format}}.
Aug. 2021.
\\\doi{10.5281/zenodo.1003149}

\bibitem[FM24]{fritzsche_hermes_plugin}
M.~Fritzsche, M.~Meinel.
hermes-plugin-python (Version 0.2.0).
Aug. 2024.
\\\doi{10.5281/zenodo.13168126}

\bibitem[JBC{\etalchar{+}}23]{JonesEtAl2023}
M.~B. Jones, C.~Boettiger, A.~Cabunoc~Mayes, A.~Smith, M.~Gruenpeter, V.~Lorentz, T.~Morrell, D.~Garijo, P.~Slaughter, K.~Niemeyer, Y.~Gil, M.~Fenner, K.~Nowak, M.~Hahnel, L.~Coy, A.~Allen, M.~Crosas, A.~Sands, N.~Chue~Hong, P.~Cruse, D.~S. Katz, C.~Goble, B.~Mecum, A.~{Gonzalez-Beltran}, N.~Ross.
{{CodeMeta}}: An Exchange Schema for Software Metadata ({{Version}} 3.0).
2023.
\\\url{https://w3id.org/codemeta/v3.0}

\bibitem[JHK21]{JayEtAl2021}
C.~Jay, R.~Haines, D.~S. Katz.
Software {{Must}} Be {{Recognised}} as an {{Important Output}} of {{Scholarly Research}}.
\emph{International Journal of Digital Curation} 16(1):6, Apr. 2021.
\\\doi{10.2218/ijdc.v16i1.745}

\bibitem[MDK{\etalchar{+}}24a]{hermes_0_8_0}
M.~Meinel, S.~Druskat, J.~Kelling, O.~Bertuch, O.~Knodel, D.~Pape.
hermes (Version v0.8.0).
Mar. 2024.
\\\url{https://archive.softwareheritage.org/swh:1:rev:b2033c915fa839e6043133e056706dce79f46d27;origin=https://github.com/softwarepub/hermes;visit=swh:1:snp:445289f60b81647ec811f300ccbe74026e8c8517}

\bibitem[MDK{\etalchar{+}}24b]{meinel_hermes}
M.~Meinel, S.~Druskat, J.~Kelling, O.~Bertuch, O.~Knodel, D.~Pape, S.~Kernchen.
hermes.
Aug. 2024.
\\\doi{10.5281/zenodo.13221383}
\\\url{https://hermes.software-metadata.pub/}

\bibitem[MDK{\etalchar{+}}24c]{hermes_0_8_1}
M.~Meinel, S.~Druskat, J.~Kelling, O.~Bertuch, O.~Knodel, D.~Pape, S.~Kernchen.
hermes (Version v0.8.1).
Aug. 2024.
\\\doi{10.5281/zenodo.13311079}

\bibitem[MDK{\etalchar{+}}24d]{hermes_0_8_1b1}
M.~Meinel, S.~Druskat, J.~Kelling, O.~Bertuch, O.~Knodel, D.~Pape, S.~Kernchen.
hermes (Version v0.8.1b1).
Aug. 2024.
\\\doi{10.5281/zenodo.13221384}

\bibitem[{PPA}]{PythonEntryPoints}
{Python Packaging Authority}.
Entry points specification.
\\\url{https://packaging.python.org/en/latest/specifications/entry-points/}

\bibitem[RLP{\etalchar{+}}21]{RaoEtAl2021}
A.~Rao, A.~Lawrence, A.~Pace, D.~Rodriguez, E.~J.~G. Gabancho, G.~Lastecoueres, I.~Mas{\'a}r, J.~Delgado, J.~Kuncar, J.~Gon{\c c}alves, L.~H. Nielsen, L.~Rossi, N.~Tarocco, N.~Harraudeau, R.~Ducceschi, S.~Hiltunen, S.~Kaplun, T.~Simko, X.~Meng.
Invenio {{Framework}}.
\emph{PyPI}, May 2021.
\\\url{https://pypi.org/project/invenio/}

\bibitem[SKN{F}16]{smithSoftwareCitationPrinciples2016}
A.~M. Smith, D.~S. Katz, K.~E. Niemeyer, {FORCE11 Software Citation Working Group}.
Software Citation Principles.
\emph{PeerJ Computer Science} 2(e86), 2016.
\\\doi{10.7717/peerj-cs.86}

\bibitem[{INV}]{TheInvenioRDMteam}
{The InvenioRDM team}.
{{InvenioRDM}}.
CERN \& contributors.
\\\url{https://inveniosoftware.org/products/rdm/}

\bibitem[Wie14]{wieringa_2014_design_science_methodology}
R.~J. Wieringa.
\emph{Design {{Science Methodology}} for {{Information Systems}} and {{Software Engineering}}}.
Springer, Berlin, 2014.

\end{thebibliography}
\bibliographystyle{eceasst}

\end{document}